# Robust distributed extended Kalman filter based on adaptive multi-kernel mixture maximum correntropy for non-Gaussian systems

Duc Viet Nguyen, Haiquan Zhao, *Senior Member, IEEE*, Jinhui Hu, and Xiaoli Li, *Fellow, IEEE*

*Abstract*— As one of the most advanced variants in the correntropy family, the multi-kernel correntropy criterion demonstrates superior accuracy in handling non-Gaussian noise, particularly with multimodal distributions. However, current approaches suffer from key limitations-namely, reliance on a single type of sensitive Gaussian kernel and the manual selection of free parameters. To address these issues and further boost robustness, this paper introduces the concept of multi-kernel mixture correntropy (MKMC), along with its key properties. MKMC employs a flexible kernel function composed of a mixture of two Student's *t*-Cauchy functions with adjustable (non-zero) means. Building on this criterion within multi-sensor networks, we propose a robust distributed extended Kalman filter-AMKMMC-RDEKF based on adaptive multi-kernel mixture maximum correntropy. To reduce communication overhead, a consensus averaging strategy is incorporated. Furthermore, an adaptive mechanism is introduced to mitigate the impact of manually tuned free parameters. At the same time, the computational complexity and convergence ability of the proposed algorithm are analyzed. The effectiveness of the proposed algorithm is validated through challenging scenarios involving power system and land vehicle state estimation.

*Index Terms*- state estimation, multi-kernel mixture correntropy, distributed extended Kalman filter.

## I. INTRODUCTION

AS a measure to improve the stability in control and monitoring systems, state estimation algorithms act as observers. Considering the problems affecting accuracy in non-Gaussian systems, it is evident that noise is a prominent and urgent issue that needs to be addressed [1,2]. First of all, we need to understand the origin of noise in nonlinear systems. In microgrids, converter-induced harmonics and switching noise are the primary noise sources, with mechanical equipment being the primary contributor to background noise [3]. In an electric vehicle, the predominant sources of noise emissions are the motor drive system and the power electronic converters [4]. In motors, the operation of squirrel-cage and synchronous induction motors produces "tooth ripple" harmonic noise, a consequence of the electromagnetic forces generated by their slotted stator and rotor assemblies [5]. In power systems, noise sources are systematically classified based on the nature of the associated physical phenomenon, including thermal, acoustic, electrical, combustion, hydrodynamic, and optical origins [6].

Furthermore, to address the impact of noise on systems, we need to know their characteristics [7,8]. In general, noise is classified based on statistical properties. For simplicity, current research often categorizes noise into two types: Gaussian noise and non-Gaussian noise. In which, addressing the impact of non-Gaussian noise, which is similar to reality, is the goal. Initially, non-Gaussian noise types were introduced with distribution properties such as the Laplace, Cauchy, and Poisson distributions [9,10,11]. However, considering the increasing requirements for accuracy and stability in modern applications, estimation algorithms need to achieve the ability to overcome non-Gaussian noise types such as impulse noise, heavy-tailed noise, and large outliers [7,8,12].

Reviewing the existing literature, state estimation is dominated by model-based (e.g., Kalman filters) and data-driven (e.g., neural networks) methods, which are the two prominent approaches today, each with distinct trade-offs [12,13,14]. Model-based approaches are computationally efficient but model-sensitive, while data-driven methods are model-agnostic but model-dependent. A promising hybrid research direction, combining Kalman filters with neural networks, is emerging to synergize their respective advantages [15,16]. It is evident that the Kalman filter, a recursive algorithm, has consistently garnered special attention due to its simplicity and efficiency. However, being developed based on the assumption of Gaussian noise is a major limitation of KF, which makes it unsuitable in current practical applications [12]. To overcome this challenge, various methods have been explored, including extending the Gaussian framework to more general stable alpha distributions, utilizing Student's *t*-distribution to handle heavy-tailed noise, and expanding the state space through Kronecker exponential modeling of both the system state and the output [17,18]. However, these studies have not yet fully addressed the influence of complex non-Gaussian noises.

Correntropy, a similarity measure, which is presented as the expected value of a kernel function between two random variables, has proven effective in signal processing, particularly in handling large outliers and non-Gaussian noise [19,20]. However, its standard implementation using a single kernel function limits the flexibility of the maximum correntropy criterion. To address this, the concept of mixture correntropy (MC), a linear combination of multiple kernels,

This work was partially supported by the National Natural Science Foundation of China (grant: 62171388, 61871461, 61571374) and Major Special Project of China Railway Group Limited (Source -2025- Special Project -001) (*Corresponding author*: Haiquan Zhao).

Duc Viet Nguyen (e-mail: ducvietpkkq@gmail.com); Jinhui Hu (e-mail: jhhu_swjtu@126.com) and Haiquan Zhao (e-mail: hqzhao_swjtu@126.com) are with the Key Laboratory of Magnetic Suspension Technology and Maglev Vehicle, Ministry of Education, School of Electrical Engineering, Southwest Jiaotong University, Chengdu, China.

Xiaoli Li (e-mail: xiaoli_li@sutd.edu.sg) is with the Singapore University of Technology and Design.

was introduced to enhance adaptability [21,22]. Despite this advancement, both the original correntropy and MC commonly rely on Gaussian kernels with fixed zero mean, which are highly sensitive and prone to producing singular matrices under certain conditions [23,24,25,26,27]. This fixed-mean assumption hampers their robustness when dealing with heavy-tailed or multimodal noise distributions. To overcome these limitations, the multi-kernel correntropy (MKC) framework was proposed, incorporating kernels with non-zero mean values for greater modeling flexibility [28,29]. More recently, integrating diverse kernel types within a unified criterion has emerged as a promising direction to mitigate the individual weaknesses of specific kernels [30,31].

Despite notable advances in handling non-Gaussian noise, the optimality of criteria within the correntropy family remains constrained by the choice of kernel bandwidth coefficients, which critically affect estimation accuracy [12,22]. In addition, the central value of the kernel function, along with the coefficients that distinguish MKC from MC, also requires careful selection [28]. These free parameters are typically set manually through extensive empirical testing [32], a process valid only under specific scenarios and targets, limiting their general applicability and optimality.

Additionally, distributed state estimation (DSE), developed through multi-sensor networks, is widely utilized to improve the reliability of control and monitoring systems [33,34]. Among various approaches, distributed Kalman filters (DKF) have emerged as reliable and secure solutions in DSE due to their scalability, fault tolerance, and cost-effectiveness [35,36]. This reliability stems from DKF's foundation in distributed processing and inter-sensor communication, where consensus algorithms ensure that each sensor gradually converges to a uniform global state estimate, even when starting with only local data. Existing consensus strategies are generally divided into three main categories: consensus on measurement (CM), consensus on estimation (CE), and consensus on information (CI) [37,38,39]. Among them, the CI method, based on iterative updates of inverse covariance matrices, stands out for its ability to maintain stability with relatively few iterations. Additionally, the development of hybrid consensus algorithms has emerged as a promising direction for enhancing performance and flexibility.

Considering the challenges arising from noise in practice and addressing the limitations in current research, this paper proposes the following contributions:

1) Aiming to address non-Gaussian noise with heavy-tailed distribution, this paper proposes a novel concept multi-kernel mixture correntropy (MKMC), which employs two types of Student's $t$-Cauchy kernels characterized by heavy-tailed behavior and flexible means (not fixed at zero). The key properties of MKMC are also analyzed.

2) Building on MKMC, a distributed extended Kalman filter (DEKF) is developed for nonlinear systems, based on multi-kernel mixture maximum correntropy (MKMMC), referred to as MKMMC-DEKF.

3) A new adaptive mechanism is designed to address the challenge of selecting free coefficients. Additionally, a thresholding strategy is introduced for measurement data to improve numerical stability.

4) Integrating the above contributions, a robust DEKF based on adaptive MKMMC (AMKMMC-RDEKF) is proposed. At the same time, the computational complexity and convergence ability of the proposed algorithm are analyzed. Its performance is verified for power system and land vehicle state estimation in complex scenarios.

This paper is organized into five parts: Section II introduces multi-kernel mixture correntropy and its properties. Section III details the construction of the MKMMC-EKF algorithm. Section IV describes the proposed algorithm. Section V provides convergence proof and computational complexity. Section VI reports the simulation results. Section VII concludes the paper.

## II. MULTI-KERNEL MIXTURE CORRENTROPY AND ITS IMPORTANT PROPERTIES

### A. Multi-kernel mixture correntropy

First, the MC criterion [21], which is based on merging two Gaussian kernels with different bandwidths, is illustrated as:

$$V_{MC}(X,Y) = \mathbf{E}\left[\theta G_{\sigma_1}(X,Y) + (1-\theta)G_{\sigma_2}(X,Y)\right] \quad (1)$$

where $X$ and $Y$ are random variables; $\sigma_1$ and $\sigma_2$ are the bandwidths of Gaussian kernel functions $G_{\sigma_1}$ and $G_{\sigma_2}$, respectively; $\theta$ is the mixture coefficient $(0 \leq \theta \leq 1)$.

However, an observable limitation of MC is that it only uses zero-mean kernels, which makes it difficult to deal with multimodal distribution noise [28]. To solve this problem, the concept of MKC was presented [28, 29], which uses kernels with non-zero mean values:

$$V_{MKC}(X,Y) = \mathbf{E}\left[\theta G_{\sigma_1}(e-a_1) + (1-\theta)G_{\sigma_2}(e-a_2)\right] \quad (2)$$

where $e = X - Y$ is an error variable; $a_1$ and $a_2$ are the center values.

**Remark 1**: It can be easily observed that MKC will become MC when $a_1 = a_2 = 0$.

To address the limitations of individual kernel types and enhance measurement performance, the concept of MKMC is introduced, where the kernel function is defined as a mixture of Student's t and Cauchy distributions:

$$V_{MKMC}(X,Y) = \mathbf{E}\left[\theta S_\alpha(e-a_1) + (1-\theta)C_\omega(e-a_2)\right] \quad (3)$$

with
$$S_\alpha(e-a_1) = \left(1 + (e-a_1)^2/(\lambda\alpha^2)\right)^{\frac{\lambda+2}{2}} \quad (4)$$

$$C_\omega(e-a_2) = 1\Big/\left[1 + (e-a_2)^2/\omega\right] \quad (5)$$

where $S_\alpha$ is Student's $t$ kernel function with bandwidth $\alpha$ [25]; $C_\omega$ is the Cauchy kernel function with bandwidth $\omega$ [24].

**Remark 2**: The proposed MKMC criterion integrates

Student's t and Cauchy kernels, both exhibiting heavy-tailed characteristics that inherently suppress impulsive outliers. Crucially, by allowing non-zero kernel centers ($a_1, a_2 \neq 0$), MKMC departs from conventional correntropy criteria (e.g., MCC, MMC), which assume symmetric error distributions centered at the origin. This design enables accurate modeling of asymmetric or multimodal error PDFs, a common occurrence in power systems under unbalanced faults, sensor biases, or mixed measurement corruption, thereby enhancing estimation robustness in practical non-Gaussian environments.

### B. Properties

In this subsection, some important properties of MKMC are analyzed and presented as follows:

**Property 1:** The kernel differences between MKMC and existing methods are illustrated in Fig. 1. Notably, this work introduces a novel kernel shape. The results show that MKMC not only inherits the advantages of MKC but also exhibits a heavy-tailed characteristic, enhancing its robustness against heavy-tailed and multimodal noise. In contrast, criteria such as correntropy, MC, and MKC based on Gaussian kernels rapidly decay to zero.

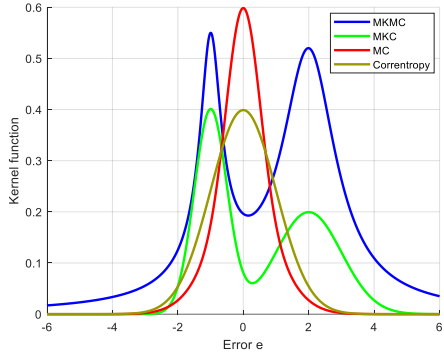

**Fig 1.** Kernel functions of correntropy, MC, MKC, and MKMC

In addition, to further clarify the difference between the MKMC introduced in this study and other hybrid kernels, Fig. 2 illustrates this. Here, the bandwidth factor of all kernels is set to 0.5, and the center value is 1. It can be easily observed that the MKMC introduced is based on the mixture of two Student's $t$ - Cauchy kernels, which have sharper and higher peaks, which is different from the mixture of Gaussian - Cauchy and Gaussian - Student's $t$. It should be noted here that studies [24,25] have demonstrated and confirmed that Student t and Cauchy kernels achieve better handling of heavy-tailed noise than Gaussian kernels. However, these studies still have limitations, including that the mean value of the kernels is always fixed to zero and the bandwidth value is also fixed, which causes them to degrade performance during dynamic estimation.

Additionally, the combination of two types of kernels will also give better performance than either kernel alone. Considering Figures 1 and 2 together, it can be seen that the proposed MKMC is more suitable for dealing with heavy-tailed noise and large outliers than existing studies.

**Property 2:** If $\theta = 1$, then MKMC is simplified to a Student's $t$ kernel with non-zero mean; Conversely, if $\theta = 0$ then it simplifies to a Cauchy kernel with non-zero mean. Furthermore, the shape of MKMC is directly affected by coefficients including center values ($a_1, a_2$), bandwidths ($\alpha, \omega$) and mixture coefficient ($\theta$), which is illustrated in Figures 3, 4, and 5, respectively.

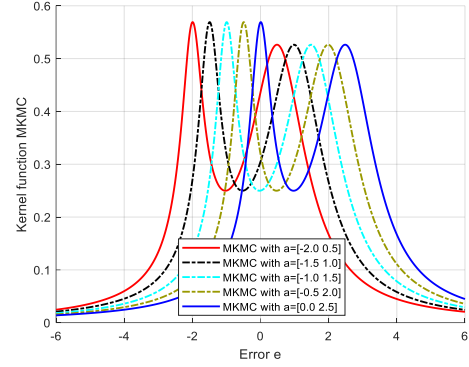

**Fig 3.** The influence of central values

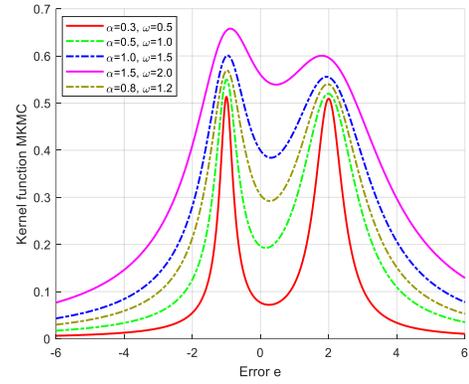

**Fig 4.** The influence of bandwidth values

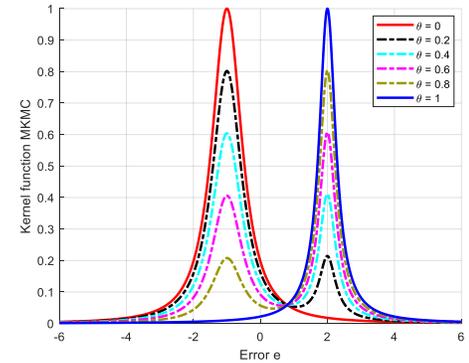

**Fig 5.** The influence of the mixture coefficient

**Property 3:** The MKMC is symmetrical, i.e, $V_{MKMC}[Y, X] = V_{MKMC}[X, Y]$.

**Property 4:** The $V_{MKMC}[X, Y]$ is positive and bounded, i.e: $0 < V_{MKMC}[X, Y] \leq 1$.

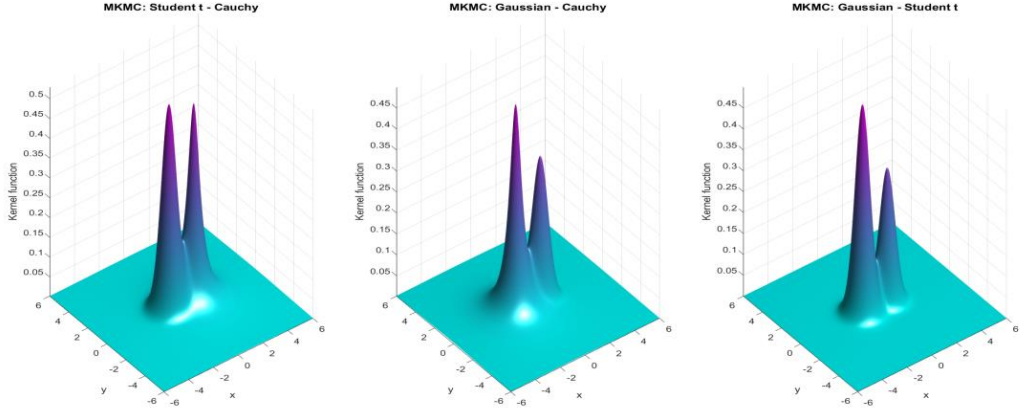

**Fig. 2.** Compare MKMC variants

***Proof***: Proof of positivity:

It can be seen that $C_\omega\left(e - a_2\right) = 1/\left[1 + \left(e - a_2\right)^2 / \omega\right] > 0$ and

$S_\alpha\left(e - a_1\right) = \left(1 + \left(e - a_1\right)^2 / \left(\lambda\alpha^2\right)\right)^{-\frac{\lambda+2}{2}} > 0$, on the other hand

$0 \leq \theta \leq 1$ therefore $V_{MKMC}\left[X, Y\right] > 0$.

Proof of bounded: It can be seen that

$S_\alpha\left(e - a_1\right) = \left(1 + \left(e - a_1\right)^2 / \left(\lambda\alpha^2\right)\right)^{-\frac{\lambda+2}{2}} \leq 1$ and

$C_\omega\left(e - a_2\right) = 1/\left[1 + \left(e - a_2\right)^2 / \omega\right] \leq \frac{1}{1} = 1$, on the other hand

$0 \leq \theta \leq 1$ therefore $V_{MKMC}\left[X, Y\right] \leq \theta*1 + (1-\theta)*1 = 1$.

Combining the two proof steps above gives: $0 < V_{MKMC}\left[X, Y\right] \leq 1$. This completes the proof.

**Property 5:** When $\alpha, \omega$ are sufficiently large and $a_1, a_2$ are sufficiently small (i.e., small error $e$), the MKMC can be approximated by:

$$V_{MKMC}\left(X, Y\right) \approx 1 - \left(\frac{\theta(\lambda+1)}{2\lambda\alpha^2} + \frac{(1-\theta)}{\omega}\right)e^2 \quad (6)$$

***Proof***: Apply the Taylor approximation to the Student's $t$ kernel to get: $S_\alpha\left(e - a_1\right) \approx 1 - \frac{\lambda+1}{2\lambda\alpha^2}e^2$, continue applying the approximation $(1+x)^n \approx 1 + nx$ to the Cauchy kernel to get: $C_\omega\left(e - a_2\right) \approx 1 - \frac{1}{\omega}e^2$, received:

$$\begin{aligned} V_{MKMC}\left(X, Y\right) &= \theta S_\alpha\left(e - a_1\right) + (1-\theta)C_\omega\left(e - a_2\right) \\ &\approx \theta\left[1 - \frac{\lambda+1}{2\lambda\alpha^2}e^2\right] + (1-\theta)\left[1 - \frac{1}{\omega}e^2\right] \\ &= 1 - \left(\frac{\theta(\lambda+1)}{2\lambda\alpha^2} + \frac{(1-\theta)}{\omega}\right)e^2 \end{aligned} \quad (7)$$

This completes the proof.

***Remark 3***: While hybrid kernel constructions (e.g., Gaussian–Cauchy [30]) have been explored in prior correntropy-based filters, the proposed MKMC is fundamentally distinct in three aspects: (i) it employs Student's $t$ and Cauchy kernels, which exhibit polynomial decay (vs. exponential decay of Gaussian-based hybrids), offering inherent robustness to heavy-tailed outliers; (ii) it allows non-zero kernel centers, enabling accurate modeling of asymmetric or multimodal error PDFs-a capability absent in zero-mean hybrids; and (iii) it integrates an adaptive mechanism that dynamically aligns kernel centers and bandwidths with the empirical error distribution, eliminating heuristic tuning. For example, at an outlier error $e = 5$, the Gaussian-Cauchy hybrid yields a kernel value of approximately 0.0061, while the proposed Student's $t$-Cauchy mixture gives 0.0080, a 31% higher influence. Thus, MKMC is not a heuristic kernel substitution, but a purpose-driven robustness criterion tailored for real-world non-Gaussian systems.

## III. DERIVATION OF THE MKMMC-EKF ALGORITHM

Consider a non-linear system of the form:

$$\mathbf{u}_k = f\left(\mathbf{u}_{k-1}\right) + \mathbf{q}_{k-1} \quad (8)$$

$$\mathbf{v}_k = h\left(\mathbf{u}_k\right) + \mathbf{r}_k \quad (9)$$

where $\mathbf{u}_k \in \mathbb{R}^n$ and $\mathbf{v}_k \in \mathbb{R}^m$ are the state vector and measurement vector at time $k$, respectively; $f(\cdot)$: nonlinear state function; $h(\cdot)$: nonlinear measurement function; $\mathbf{r}_k$: measurement noise with covariance matrix satisfying $E\left[\mathbf{r}_k \mathbf{r}_k^T\right] = \mathbf{R}_k$; $\mathbf{q}_{k-1}$: process noise with covariance matrix satisfying $E\left[\mathbf{q}_{k-1}\mathbf{q}_{k-1}^T\right] = \mathbf{Q}_{k-1}$.

Similar to other extended Kalman filters [40], the MKMMC-EKF algorithm includes two steps:

*1) **Prediction step***: The prior state mean $\hat{\mathbf{u}}_k$ and the corresponding covariance matrix $\hat{\mathbf{P}}_k$ are calculated as:

$$\hat{\mathbf{u}}_k = f\left(\hat{\mathbf{u}}_{k-1}\right) \quad (10)$$

$$\hat{\mathbf{P}}_k = \mathbf{G}_{k-1}\hat{\mathbf{P}}_{k-1}\mathbf{G}_{k-1}^T + \mathbf{Q}_{k-1} \quad (11)$$

where $\mathbf{G}_{k-1}$ is the Jacobian matrix and $\mathbf{G}_{k-1} = \frac{\partial f}{\partial \mathbf{u}}\bigg|_{\mathbf{u} = \hat{\mathbf{u}}_{k-1}}$.

*2) Update step*: First of all, the measurement function (9) is approximated by [12]:

$$\mathbf{v}_k \approx h(\hat{\mathbf{u}}_k) + \mathbf{H}_k (\mathbf{u}_k - \hat{\mathbf{u}}_k) + \mathbf{r}_k \quad (12)$$

where $\mathbf{H}_k$ is the Jacobian matrix and $\mathbf{H}_k = \dfrac{\partial h}{\partial \mathbf{u}}\Big|_{\mathbf{u} = \hat{\mathbf{u}}_k}$.

Combining equations (8), (9), (10), and (11) yields a non-linear model as:

$$\begin{bmatrix} \hat{\mathbf{u}}_k \\ \mathbf{v}_k - h(\hat{\mathbf{u}}_k) + \mathbf{H}_k \hat{\mathbf{u}}_k \end{bmatrix} = \begin{bmatrix} \mathbf{I} \\ \mathbf{H}_k \end{bmatrix} \mathbf{u}_k + \Delta_k \quad (13)$$

where $\mathbf{I}$ is an identity matrix; $\Delta_k = \begin{bmatrix} -(\mathbf{u}_k - \hat{\mathbf{u}}_k) \\ \mathbf{r}_k \end{bmatrix}$.

The covariance matrix of $\Delta_k$ can be described as:

$$E\left[\Delta_k \Delta_k^T\right] = \begin{bmatrix} \mathbf{P}_k & 0 \\ 0 & \mathbf{R}_k \end{bmatrix} = \begin{bmatrix} \mathbf{C}_{k,\mathbf{P}} \mathbf{C}_{k,\mathbf{P}}^T & 0 \\ 0 & \mathbf{C}_{k,\mathbf{R}} \mathbf{C}_{k,\mathbf{R}}^T \end{bmatrix} = \mathbf{C}_k \mathbf{C}_k^T \quad (14)$$

where $\mathbf{C}_k$ is the result obtained after performing the Cholesky decomposition. Note that the conditions for performing the Cholesky decomposition are assumed to be satisfied.

Multiply both sides of Eq.(13) by $\mathbf{C}_k^{-1}$, obtaining the linear regression problem:

$$\mathbf{z}_k = \mathbf{W}_k \mathbf{u}_k + \mathbf{e}_k \quad (15)$$

where

$$\mathbf{z}_k = \mathbf{C}_k^{-1} \begin{bmatrix} \hat{\mathbf{u}}_k \\ \mathbf{v}_k - h(\hat{\mathbf{u}}_k) + \mathbf{H}_k \hat{\mathbf{u}}_k \end{bmatrix} \quad (16)$$

$$\mathbf{W}_k = \mathbf{C}_k^{-1} \begin{bmatrix} \mathbf{I} \\ \mathbf{H}_k \end{bmatrix} \quad (17)$$

$$\mathbf{e}_k = \mathbf{C}_k^{-1} \Delta_k \quad (18)$$

Then, the cost function based on MKMC is defined as

$$J_{MKMC}(\mathbf{u}_k) = \frac{1}{N} \sum_{i=1}^{N} \left[ \theta S_\alpha (e_{k,i} - a_1) + (1-\theta) C_\omega (e_{k,i} - a_2) \right] \quad (19)$$

where $e_{k,i} = z_{k,i} - \mathbf{w}_{k,i} \mathbf{u}_{k,i}$, and $N = n + m$.

The optimal estimate of $\hat{\mathbf{u}}_k$ -based MKMC can be calculated as follows:

$$\hat{\mathbf{u}}_k = \arg\max_{\mathbf{u}_k} J_{MKMC}(\mathbf{u}_k) \quad (20)$$

Accordingly, the result of $\hat{\mathbf{u}}_k$ can be found by

$$\frac{\partial J_{MKMC}(\mathbf{u}_k)}{\partial \mathbf{u}_k} = 0 \quad (21)$$

The achieved results can be expressed as follows:

$$\mathbf{u}_k = \left( \mathbf{W}_k^T \mathbf{D}_k \mathbf{W}_k \right)^{-1} \mathbf{W}_k^T \mathbf{D}_k \mathbf{z}_k \quad (22)$$

where

$$\mathbf{D}_k = \begin{bmatrix} \mathbf{D}_{k,\mathbf{u}} & 0 \\ 0 & \mathbf{D}_{k,\mathbf{v}} \end{bmatrix} \quad (23)$$

$$\mathbf{D}_{k,\mathbf{u}} = diag\left( \Theta_{k,1}(e_{k,1}),...,\Theta_{k,n}(e_{k,n}) \right) \quad (24)$$

$$\mathbf{D}_{k,\mathbf{v}} = diag\left( \Theta_{k,n+1}(e_{k,n+1}),...,\Theta_{k,n+m}(e_{k,n+m}) \right) \quad (25)$$

$$\Theta_{k,i}(e_{k,i}) = \theta S_\alpha (e_{k,i} - a_1) + (1-\theta) C_\omega (e_{k,i} - a_2) \quad (26)$$

This estimated result weights each error component $e_{k,i}$ by $\Theta_{k,i}$, which is large when the error is small and decays polynomially for large errors. Unlike Gaussian-based correntropy that suppresses outliers exponentially, the Student's $t$–Cauchy mixture ensures that heavy-tailed errors still contribute meaningfully to the estimation, enhancing robustness without sacrificing accuracy. It can be easily observed that the optimal estimate value $\hat{\mathbf{u}}_k$ can be found based on the fixed point iteration method, i.e.:

$$\hat{\mathbf{u}}_k^{(t)} = g\left(\hat{\mathbf{u}}_k^{(t-1)}\right) = \left( \mathbf{W}_k^T \mathbf{D}_k^{(t-1)} \mathbf{W}_k \right)^{-1} \mathbf{W}_k^T \mathbf{D}_k^{(t-1)} \mathbf{z}_k \quad (27)$$

Then, applying the matrix inversion lemma results:

$$\hat{\mathbf{u}}_k^{(t)} = \hat{\mathbf{u}}_k^{(t-1)} + \mathbf{K}_k^{(t-1)} \left( \mathbf{v}_k - h\left(\hat{\mathbf{u}}_k^{(t-1)}\right) \right) \quad (28)$$

where

$$\mathbf{K}_k^{(t-1)} = \hat{\mathbf{P}}_k^{(t-1)} \mathbf{H}_k^T \left( \mathbf{H}_k \hat{\mathbf{P}}_k^{(t-1)} \mathbf{H}_k^T + \mathbf{R}_k^{(t-1)} \right)^{-1} \quad (29)$$

$$\hat{\mathbf{P}}_k^{(t-1)} = \mathbf{C}_{k,\mathbf{P}} \left( \mathbf{D}_{k,\mathbf{u}}^{(t-1)} \right)^{-1} \mathbf{C}_{k,\mathbf{P}}^{(t-1)} \quad (30)$$

$$\mathbf{R}_k^{(t-1)} = \mathbf{C}_{k,\mathbf{R}} \left( \mathbf{D}_{k,\mathbf{v}}^{(t-1)} \right)^{-1} \mathbf{C}_{k,\mathbf{R}}^{(t-1)} \quad (31)$$

Additionally, the covariance matrix can be updated by:

$$\hat{\mathbf{P}}_k^{(t)} = \left( \mathbf{I} - \mathbf{K}_k^{(t-1)} \mathbf{H}_k \right) \hat{\mathbf{P}}_k^{(t-1)} \left( \mathbf{I} - \mathbf{K}_k^{(t-1)} \mathbf{H}_k \right)^{-T} + \mathbf{K}_k^{(t-1)} \mathbf{R}_k \left( \mathbf{K}_k^{(t-1)} \right)^T \quad (32)$$

Finally, the steps of MKMMC-EKF are summarized as **Algorithm 1**.

---

**Algorithm 1: MKMMC-EKF**

---

1: **Initialization:** $k = 1$ ; $\hat{\mathbf{u}}_0$ ; $\hat{\mathbf{P}}_0$ ; $\alpha$ ; $\omega$ ; $\theta$ ; $\lambda$ ; $\varepsilon$ (threshold)

2: **Predict:** $\hat{\mathbf{u}}_k = f\left(\hat{\mathbf{u}}_{k-1}\right)$ and $\hat{\mathbf{P}}_k = \mathbf{G}_{k-1} \hat{\mathbf{P}}_{k-1} \mathbf{G}_{k-1}^T + \mathbf{Q}_{k-1}$

3: **Update:** $t = 1$ and $\hat{\mathbf{u}}_k^{(t)} = \hat{\mathbf{u}}_k^{(t-1)}$

   Iteration: $\hat{\mathbf{u}}_k^{(t)} = \hat{\mathbf{u}}_k^{(t-1)} + \mathbf{K}_k^{(t-1)} \left( \mathbf{v}_k - h\left(\hat{\mathbf{u}}_k^{(t-1)}\right) \right)$

   $\mathbf{K}_k^{(t-1)} = \hat{\mathbf{P}}_k^{(t-1)} \mathbf{H}_k^T \left( \mathbf{H}_k \hat{\mathbf{P}}_k^{(t-1)} \mathbf{H}_k^T + \mathbf{R}_k^{(t-1)} \right)^{-1}$

   $\hat{\mathbf{P}}_k^{(t-1)} = \mathbf{C}_{k,\mathbf{P}} \left( \mathbf{D}_{k,\mathbf{u}}^{(t-1)} \right)^{-1} \mathbf{C}_{k,\mathbf{P}}^{(t-1)}$

   $\mathbf{R}_k^{(t-1)} = \mathbf{C}_{k,\mathbf{R}} \left( \mathbf{D}_{k,\mathbf{v}}^{(t-1)} \right)^{-1} \mathbf{C}_{k,\mathbf{R}}^{(t-1)}$

   $e_{k,i} = z_{k,i} - \mathbf{w}_{k,i} \hat{\mathbf{u}}_{k,i}^{(t-1)}$

   $\hat{\mathbf{P}}_k^{(t)} = \left( \mathbf{I} - \mathbf{K}_k^{(t-1)} \mathbf{H}_k \right) \hat{\mathbf{P}}_k^{(t-1)} \left( \mathbf{I} - \mathbf{K}_k^{(t-1)} \mathbf{H}_k \right)^{-T} + \mathbf{K}_k^{(t-1)} \mathbf{R}_k \left( \mathbf{K}_k^{(t-1)} \right)^T$

   **if** $\left\| \hat{\mathbf{u}}_k^{(t)} - \hat{\mathbf{u}}_k^{(t-1)} \right\| / \left\| \hat{\mathbf{u}}_k^{(t-1)} \right\| \leq \varepsilon$ **then** go to step 4

   **else** $t = t + 1$ , go to step iteration

   **end**

4: $k = k + 1$ , go to step 2

---

## IV. Proposed Algorithm AMKMMC-RDEKF

It can be observed that the previously presented MKMMC-DEKF faces several limitations, including communication overhead, measurement values exceeding predefined thresholds, and the manual selection of free coefficients. To address these issues, the following solutions are proposed:

### A. Consensus average algorithm

To effectively reduce the communication burden in sensor networks, a consensus averaging algorithm is implemented to fuse information from all sensors. Specifically, with a network consisting of $b$ sensors, the communication topology can be described through an undirected connection graph $\psi = (A, B)$, in which $A = \{1, 2, ..., b\}$ represents sensor nodes, $B \subseteq A \times A$ represents connections. The neighbors of the $i$-th sensor are $N_i = \{j \in A : (i, j) \in B\} \cup \{i\}$, that is, the neighbors consist itself. Assuming the initial value $\psi(0)$, the consensus average value of the $i$-th sensor can be calculated by:

$$\psi_i(l) = \sum_{j \in N_i} \pi_{i,j} \psi_j(l-1) \tag{33}$$

where $\pi_{i,j}$ is the weight coefficient, in which $\sum_{j \in N_i} \pi_{i,j} = 1$; $l = 1, 2, ..., L$ is the number of iterations.

Furthermore, Eq.(33) satisfies the condition:

$$\lim_{L \to \infty} \psi_i(L) = \frac{1}{b} \sum_{i=1}^{b} \psi_i(0) \tag{34}$$

**Remark 4**: The integration of consensus averaging not only reduces communication overhead but also enhances resilience to node failures. Unlike centralized estimators, AMKMMC-RDEKF maintains estimation accuracy even under partial network disconnections, a critical requirement for modern, highly distributed power grids.

### B. Determine the free coefficients

The impact of manually selected bandwidth values in kernel functions on the performance of state estimation algorithms has been analyzed in [21,22]. These coefficients play a critical role in handling outliers and non-Gaussian noise. Although [32] acknowledges this issue, it only provides suggested value ranges rather than a generalizable solution. Therefore, a robust and adaptive approach applicable across various systems and environmental conditions is required. To begin, formula (3) can be rewritten as:

$$U_{0,\eta,a} = E\left[\sum_{j=1}^{2} \theta_j \kappa_{\eta_j}(e - a_j)\right]$$
$$= \frac{1}{2} \int_{-\infty}^{\infty} \left(\sum_{j=1}^{2} \theta_j \kappa_{\eta_j}(e - a_j)\right)^2 dv + \frac{1}{2} \int_{-\infty}^{\infty} (\rho_e(v))^2 dv \tag{35}$$
$$- \frac{1}{2} \int_{-\infty}^{\infty} \left(\sum_{j=1}^{2} \theta_j \kappa_{\eta_j}(e - a_j) - \rho_e(v)\right)^2 dv$$

where $\eta_1 = \alpha$; $\eta_2 = \omega$; $\rho_e(v)$ is the PDF of the error variable.

Considering the real-world object model, $U_{0,\eta,a}$ can be simplified to:

$$V_{0,\eta,a} = \frac{1}{2} \int_{-\infty}^{\infty} (\rho_e(v))^2 dv$$
$$- \frac{1}{2} \int_{-\infty}^{\infty} \left(\sum_{j=1}^{2} \theta_j \kappa_{\eta_j}(e - a_j) - \rho_e(v)\right)^2 dv \tag{36}$$

Suppose $\Psi_\theta$, $\Psi_\eta$, and $\Psi_a$ are the sets of values of the coefficient vectors $\theta$, $\eta$, and $a$, respectively. Then the optimization process is set up as follows:

$$\left(\theta^*, \eta^*, a^*\right) = \arg\max_{\theta \in \Psi_\theta, \eta \in \Psi_\eta, a \in \Psi_a} V_{0,\eta,a}$$
$$= \arg\max_{\theta \in \Psi_\theta, \eta \in \Psi_\eta, a \in \Psi_a} -\frac{1}{2} \int_{-\infty}^{\infty} \left(\sum_{j=1}^{2} \theta_j \kappa_{\eta_j}(e - a_j)\right)^2 dv$$
$$+ E\left(\sum_{j=1}^{2} \theta_j \kappa_{\eta_j}(e - a_j)\right) \tag{37}$$
$$= \arg\max_{\theta \in \Psi_\theta, \eta \in \Psi_\eta, a \in \Psi_a} -\frac{1}{2} \theta^T \left(\int_{-\infty}^{\infty} \tilde{\mathbf{f}}(v) \tilde{\mathbf{f}}^T(v) dv\right) \theta + \theta^T \zeta$$
$$= \arg\max_{\theta \in \Psi_\theta, \eta \in \Psi_\eta, a \in \Psi_a} -\frac{1}{2} \theta^T \Theta \theta + \theta^T \Gamma$$

where $\tilde{\mathbf{f}}(v) = \left[\kappa_{\eta_1}(e - a_1), \kappa_{\eta_2}(e - a_2)\right]^T$, $\Gamma = \frac{1}{N} \sum_{i=1}^{N} \tilde{\mathbf{f}}(v_i)$, and $\Theta = \int_{-\infty}^{\infty} \tilde{\mathbf{f}}(v) \tilde{\mathbf{f}}^T(v) dv$.

The free coefficients can be found by maximizing the objective function:

$$\tilde{V} = -\frac{1}{2} \theta^T \Theta \theta + \theta^T \Gamma \tag{38}$$

The mixture coefficient $\theta$ is found by maximizing:

$$\theta^* = (\Theta + \gamma \mathbf{I})^{-1} \Gamma \tag{39}$$

where $\gamma$ is the regularization coefficient.

With the above guidelines, the free coefficients, which are found at time $k$, are summarized in **Algorithm 2**.

---

**Algorithm 2: Determine the free coefficients**

1: Initialization: $\gamma$, $\Psi_\eta$, $\tilde{\mathbf{f}}(v)$, $\Gamma$, and $\Theta$

2: Determine the center values:
By applying K-means clustering to determine the center vector $a^*$ over the error samples $\{e_i(k)\}_{i=1}^{N}$

3: Determine the kernel sizes:
for all i=1,2,…, b do
  for all i=1,2 do
$$\eta_j^* = \arg\max_{\eta \in \Psi_\eta} \left[ -\left((\Theta + \gamma \mathbf{I})^{-1} \Gamma\right)^T \mathbf{H}\left((\Theta + \gamma \mathbf{I})^{-1} \Gamma\right) \right]$$
$$+ 2\left((\Theta + \gamma \mathbf{I})^{-1} \Gamma\right)^T \Gamma$$
  end for
end for

4: Determine the mixture coefficient:
Calculate $\theta^* = (\Theta + \gamma \mathbf{I})^{-1} \Gamma$

**Remark 5**: The proposed adaptive mechanism is grounded in statistical learning: K-means identifies dominant error modes as kernel centers, while the closed-form solution (39) minimizes the Euclidean distance between the kernel mixture and the empirical error PDF. This eliminates manual tuning and ensures the MKMC criterion adapts to time-varying noise structures in power systems or vehicle tracking.

### C. Measurement values exceeding the threshold

In practice, measured values may occasionally exceed the threshold (i.e., become extremely large), leading to systematic errors due to the near-singularity of $\mathbf{D}_k$. To mitigate this issue, the following approach is designed:

$$\Pi_k = \mathbf{v}_k - h\left(\hat{\mathbf{u}}_k^{(t/t-1)}\right) \tag{40}$$

$$\vartheta_k = \mathbf{H}_k \hat{\mathbf{P}}_k^{(t/t-1)} \mathbf{H}_k^T \tag{41}$$

$$\wp_k = \Pi_k^T \vartheta_k^{-1} \Pi_k \tag{42}$$

Specifically, this solution is implemented as follows: if $\left|\wp_k\right| > \mu$ ( $\mu$ is a preset threshold) then $\hat{\mathbf{u}}_k^{(t)} = \hat{\mathbf{u}}_k^{(t-1)}$ and $\hat{\mathbf{P}}_k^{(t)} = \hat{\mathbf{P}}_k^{(t-1)}$. Otherwise, if $\left|\wp_k\right| \le \mu$, all update and prediction steps are performed as usual.

### D. Derivation of the proposed algorithm

In this section, the AMKMMC-RDEKF algorithm is developed for sensor networks, incorporating a consensus averaging algorithm to alleviate communication burden. In parallel, adaptive determination of free coefficients and the establishment of a measurement data threshold are implemented to enhance accuracy and numerical stability.

First, based on the Appendix A provided in this study, the estimated value $\hat{\mathbf{u}}_k^{(t/t)}$ is expressed as an estimate of a single node (note that the information from neighboring nodes is not incorporated at this stage):

$$\hat{\mathbf{u}}_k^{(t/t)} = \hat{\mathbf{u}}_k^{(t/t-1)} + \hat{\mathbf{P}}_k^{(t/t)}\left[\hat{\mathbf{v}}_k^{(t/t)} - h\left(\hat{\mathbf{u}}_k^{(t/t-1)}\right) + \mathbf{H}_k^{(t/t)}\hat{\mathbf{u}}_k^{(t/t-1)}\right] - \mathbf{H}_k^{(t/t)}\hat{\mathbf{u}}_k^{(t/t-1)}$$

$$= \hat{\mathbf{P}}_k^{(t/t)}\left[\begin{matrix}\left(\hat{\mathbf{P}}_k^{(t/t-1)}\right)^{-1}\hat{\mathbf{u}}_k^{(t/t-1)} + \\ \sum_{i=1}^{b}\left(\mathbf{H}_k^{(t/t)}\right)^T \mathbf{R}_k\left(\hat{\mathbf{v}}_k^{(t/t)} - h\left(\hat{\mathbf{u}}_k^{(t/t-1)}\right) + \mathbf{H}_k^{(t/t)}\hat{\mathbf{u}}_k^{(t/t-1)}\right)\end{matrix}\right] \tag{43}$$

Next, the consensus averaging algorithm is devised on a sensor network as:

$$\mathbf{\Omega}_i(l) = \sum_{j\in N_i}\pi_{i,j}\mathbf{\Omega}_j(l-1) \tag{44}$$

where $\quad\mathbf{\Omega}_i(0) = \left(\mathbf{H}_i^{(t/t)}\right)^T \mathbf{R}_i \mathbf{H}_i^{(t/t)} \tag{45}$

$$\mathbf{\Xi}_i(l) = \sum_{j\in N_i}\pi_{i,j}\mathbf{\Xi}_j(l-1) \tag{46}$$

with $\mathbf{\Xi}_i(0) = \left(\mathbf{H}_i^{(t/t)}\right)^T \mathbf{R}_i\left(\hat{\mathbf{v}}_i^{(t/t)} - h_i\left(\hat{\mathbf{u}}_i^{(t/t-1)}\right) + \mathbf{H}_i^{(t/t)}\hat{\mathbf{u}}_i^{(t/t-1)}\right) \tag{47}$

The weight coefficient $\pi_{i,j}$ can be calculated via the Metropolis weights rule [41]:

$$\pi_{i,j} = \begin{cases} 1/\left(1 + \max(d_i, d_j)\right) & \text{if } \{i, j\} \in B \\ 1 - \sum_{\{i,j\}\in\psi}\pi_{i,j} & \text{if } i = j \\ 0 & \text{else} \end{cases} \tag{48}$$

where $d_i$ represents the degree of the $i$-th node, i.e., the number of connected branches.

Finally, the updated equation of the $i$-th sensor is calculated as follows:

$$\hat{\mathbf{u}}_i^{(t/t)} \approx \hat{\mathbf{P}}_i^{(t/t)}\left[\left(\hat{\mathbf{P}}_i^{(t/t-1)}\right)^{-1}\hat{\mathbf{u}}_i^{(t/t-1)} + b\mathbf{\Xi}_i(L)\right] \tag{49}$$

$$\hat{\mathbf{P}}_i^{(t/t)} \approx \left[\left(\hat{\mathbf{P}}_i^{(t/t-1)}\right)^{-1} + b\mathbf{\Omega}_i(L)\right]^{-1} \tag{50}$$

Based on the above analyses, the pseudocode of the proposed distributed estimation algorithm, AMKMMC-RDEKF, is summarized in **Algorithm 3**.

| **Algorithm 3: AMKMMC-RDEKF** |
|---|
| 1:    Consider a network with $b$ sensors. |
| 2:    Calculate $\hat{\mathbf{u}}_k^{(t)}$ and $\hat{\mathbf{P}}_k^{(t)}$ via **Algorithm 1**, in which the coefficients $\alpha$ ; $\omega$ ; and $\theta$ are found employing **Algorithm 2**, and establish the measurement data threshold. |
| 3:    Calculate $\mathbf{\Omega}_i(l) = \sum_{j\in N_i}\pi_{i,j}\mathbf{\Omega}_j(l-1)$ <br><br> $\mathbf{\Xi}_i(l) = \sum_{j\in N_i}\pi_{i,j}\mathbf{\Xi}_j(l-1)$ |
| 4:    Update $\hat{\mathbf{P}}_i^{(t/t)} \approx \left[\left(\hat{\mathbf{P}}_i^{(t/t-1)}\right)^{-1} + b\mathbf{\Omega}_i(L)\right]^{-1}$ <br><br> $\hat{\mathbf{u}}_i^{(t/t)} \approx \hat{\mathbf{P}}_i^{(t/t)}\left[\left(\hat{\mathbf{P}}_i^{(t/t-1)}\right)^{-1}\hat{\mathbf{u}}_i^{(t/t-1)} + b\mathbf{\Xi}_i(L)\right]$ <br><br> Then go to step 2, and $t \to t+1$ . |

## V. CONVERGENCE ANALYSIS AND COMPUTATIONAL COMPLEXITY

In this section, the computational complexity and convergence ability of the proposed algorithm are analyzed.

### A. Convergence analysis

In Section III, starting from Equation (22), we stated that the value $\mathbf{u}_k$ will be obtained based on the fixed-point algorithm, i.e., the iteration $\hat{\mathbf{u}}_k^{(t)} = g\left(\hat{\mathbf{u}}_k^{(t-1)}\right)$ (Eq. (27)) will converge to a unique point. This is easily verified based on the Banach fixed-point theorem, which is also known as the contraction mapping theorem.

According to the Banach fixed-point theorem, a fixed-point algorithm will be guaranteed to converge if $\exists\phi > 0$ and $1 > \Upsilon > 0$ such that if the initial weight vector $\|\mathbf{u}_0\|_p \le \phi$ , and $\forall\mathbf{u} \in \left\{\mathbf{u} \in \mathbb{R}^L : \|\mathbf{u}_0\|_p \le \phi\right\}$ , it holds that:

$$\begin{cases} \left\| \mathbf{g}(\mathbf{u}) \right\|_p \le \phi \\ \left\| \nabla_{\mathbf{u}} \mathbf{g}(\mathbf{u}) \right\|_p = \left\| \dfrac{\partial \mathbf{g}(\mathbf{u})}{\partial \mathbf{u}^T} \right\|_p \le \Upsilon \end{cases} \tag{51}$$

where $\left\| \cdot \right\|_p$ is illustrate an $\ell_p$-norm for a vector, and $\nabla_{\mathbf{u}} \mathbf{g}(\mathbf{u})$ is illustrate the $m \times m$ Jacobian matrix of $\mathbf{g}(\mathbf{u})$ with respect to $\mathbf{u}$, given by:

$$\nabla_{\mathbf{u}} \mathbf{g}(\mathbf{u}) = \begin{bmatrix} \dfrac{\partial}{\partial u_1} \mathbf{g}(\mathbf{u}) & \dfrac{\partial}{\partial u_2} \mathbf{g}(\mathbf{u}) & \dots & \dfrac{\partial}{\partial u_m} \mathbf{g}(\mathbf{u}) \end{bmatrix} \tag{52}$$

A sufficient condition for ensuring the convergence of the proposed algorithm is established through the two theorems presented below. For simplicity, the bandwidths of the two kernels are denoted as $\eta_i$ (similar to Equation 35):

***Theorem 1:*** If $\phi > \zeta = \dfrac{\Lambda \sqrt{m}}{\theta_{\min} \left( \sum_{j=1}^N \sum_{i=1}^m \dfrac{\theta_i}{\gamma_i} \Gamma_j \Gamma_i^T \right) + \theta_r}$ and

$\eta \ge \eta^*$, in which $\Lambda = \sum_{j=1}^N \sum_{i=1}^m \left( \theta_i / \gamma_i^3 \right) \left| t_j - a_i \right| \left\| \Gamma_j \right\|_1$, $\eta^*$ is the solution of the equation $\varphi(\eta) = \left( \Lambda \sqrt{m} \right) / \left( \theta_{\min}(\delta) + \theta_r \right)$, $\delta = \sum_{j=1}^N \sum_{i=1}^m \left( \theta_i / \gamma_i^3 \right) \exp \left( - \left[ \left( \left\| \Gamma_j \right\|_1 + \left| t_j - a_i \right| \right)^2 / \left( \Gamma_i \Gamma_i^T \right) \right] \right)$, and $\theta_{\min} [\cdot]$ illustrate the minimum eigenvalue of the matrix term. Then $\left\| \mathbf{g}(\mathbf{u}) \right\|_1 \le \phi$ for all $\mathbf{u} \in \left\{ \mathbf{u} \in \mathbb{R}^L : \left\| \mathbf{u} \right\|_1 \le \phi \right\}$.

The proof of Theorem 1 is provided in Appendix B.

***Theorem 2:*** If $\phi > \zeta$ and $\eta \ge \max \left( \eta^*, \eta^\dagger \right)$, in which $\eta^*$ is the solution of equation $\varphi(\eta) = \phi$, and $\eta^\dagger$ is the solution of equation $\chi(\eta) = \hbar$, $1 > \hbar > 0$, where $\eta \in (0, \infty)$ and $\chi(\eta) = \left[ \left( \sqrt{m} \Pi \right) / \left( \left( \theta_{\min}(\delta) + \gamma'' \right) \eta^2 \right) \right]$, with $\Pi = \sum_{j=1}^N \sum_{i=1}^m \dfrac{\theta_i}{\gamma_i^5} \left( \phi \left\| \Gamma_j \right\|_1 + \left| t_j - a_i \right| \right) \left\| \Gamma_j \right\|_1 \left( \phi \left\| \Gamma_j^T \Gamma_j \right\|_1 + \left| t_j \right| \left\| \Gamma_j \right\|_1 \right)$. Then, it holds that $\left\| \mathbf{g}(\mathbf{u}) \right\|_1 \le \phi$ and $\left\| \nabla_{\mathbf{u}} \mathbf{g}(\mathbf{u}) \right\|_1 \le \hbar$ for all $\forall \mathbf{u} \in \left\{ \mathbf{u} \in \mathbb{R}^L : \left\| \mathbf{u} \right\|_1 \le \phi \right\}$.

The proof of Theorem 2 is provided in Appendix C.

According to the Banach fixed-point theorem and Theorem 2, with an initial weight vector satisfying $\left\| \mathbf{u}_0 \right\|_1 \le \phi$, the convergence of the proposed algorithm to a unique fixed point within the specified range can be assured $\mathbf{u} \in \left\{ \mathbf{u} \in \mathbb{R}^L : \left\| \mathbf{u} \right\|_1 \le \phi \right\}$, in which the kernel bandwidth $\eta$ is larger than a certain value.

### B. Computational complexity

To comprehensively compare the proposed algorithm with existing algorithms, the computational complexity is provided in Table I.

TABLE I
COMPUTATIONAL COMPLEXITY

| Equation | Multiplication, addition and subtraction | Division, Cholesky decomposition, exponentiation, and matrix inversion |
|---|---|---|
| (28) | $2nm$ | 0 |
| (29) | $mn^2 + 4m^2n - m^2$ | $O(m^3)$ |
| (30) | $4n^3 - n^2$ | $n + O(n^3)$ |
| (31) | $4m^3 - m^2$ | $m + O(m^3)$ |
| (32) | $8n^3 + 4nm^2 - 2nm + 4n^2$ | 0 |

From the results presented in Table I, it is evident that the computational complexity of the proposed algorithm is comparable to that of the algorithms based on MKC and MC [21,28]. However, the differences in the kernel and the mechanism for adaptive determination of free coefficients will help them to increase the estimation accuracy.

## VI. SIMULATION RESULTS

In this section, the accuracy and stability of the proposed algorithms are evaluated and compared with existing methods, including DEKF [42], MCC-DEKF [43], and MMC-DEKF [21] for the task of power system and land vehicle state estimation under complex conditions. Specifically, simulations are conducted over a total sample time of T=100, and 100 independent Monte Carlo experiments. Simultaneously, the free coefficients employed in the compared algorithms are listed in Table II. It should be noted that these coefficients are selected based on both referenced literature and experimental tuning. To assess accuracy, the root mean square error (RMSE) is employed, defined as follows:

$$\text{RMSE}(t) = \sqrt{\frac{1}{100} \sum_{i=1}^{100} \left\| \mathbf{u}_i^{(t|t)} - \hat{\mathbf{u}}_i^{(t|t)} \right\|_2^2} \tag{53}$$

where: $\mathbf{u}_i^{(t|t)}$ and $\hat{\mathbf{u}}_i^{(t|t)}$ are the true value and estimated value of state. From here, the average RMSE (ARMSE) criterion can be derived as follows: $\text{ARMSE} = \frac{1}{300} \sum_{t=1}^{300} \text{RMSE}(t)$.

At the same time, the mean absolute error (MAE) is defined as:

$$\text{MAE}(t) = \frac{1}{300} \sum_{i=1}^{300} \left| \hat{\mathbf{u}}_i^{(t|t)} - \mathbf{u}_i^{(t|t)} \right| \tag{54}$$

TABLE II
VALUES OF COEFFICIENTS

| Coefficient / Algorithm | Center value | Threshold | Bandwidth |
|---|---|---|---|
| MCC-DEKF | ./. | $\varepsilon = 10^{-6}$ | $\sigma = 1.8$ |
| MMC-DEKF | ./. | $\varepsilon = 10^{-6}$ | $\sigma_1 = 1.6;\ \sigma_2 = 1.2$ |
| MKMMC-DEKF | $a_1 = -0.025$ $a_2 = 0.0032$ | $\varepsilon = 10^{-6}$ | $\omega = 1.5;\ \alpha = 2.2$ |

Furthermore, some of the noise models that impact the systems are illustrated as follows:

① The mixed-Gaussian noise:

$$r_{mG} \sim 0.4 * G\left(0, 10^{-1}\right) + 0.6 * G\left(0, 25\right) \tag{55}$$

② The Rayleigh noise:

$$r_{Ray}(t) \sim \frac{t}{9}\exp\left(-\frac{t^2}{18}\right) \tag{56}$$

where $G(0, 10^{-1})$ is a Gaussian distribution with mean 0.5 and variance $10^{-1}$; other Gaussian distributions are similar.

At the same time, the sensor network used in this study has the communication topology illustrated in Figure 6.

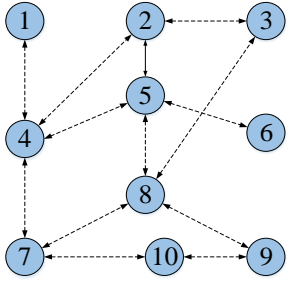

**Fig. 6.** Topology of sensor network.

### A. Power system state estimation

Consider the IEEE-14bus power system [22,23], in which the initial state $\mathbf{u}_i^{(0|0)}$ is employed as in [44], and $\mathbf{P}_i^{(0|0)} = 10^{-2}$, $\mathbf{R}_i = 10^{-2}\mathbf{I}_m$ ($m = 96$, number of measurements), $\mathbf{Q}_i = 10^{-5}\mathbf{I}_n$ ($n = 27$, number of states). Additionally, the state transition function $f(\mathbf{u})$ is derived from Holt's two-parameter exponential smoothing technique, and the measurement function $h(\mathbf{u})$ can be described by:

$$\begin{cases} P_i = \sum_{j=1}^{N} |B_i||B_j|\left(M_{ij}\cos\beta_{ij} + Z_{ij}\sin\beta_{ij}\right) \\ Q_i = \sum_{j=1}^{N} |B_i||B_j|\left(M_{ij}\sin\beta_{ij} - Z_{ij}\cos\beta_{ij}\right) \\ P_{ij} = B_i^2\left(M_{gi} + M_{ij}\right) - |B_i||B_j|\left(M_{ij}\cos\beta_{ij} + Z_{ij}\sin\beta_{ij}\right) \\ Q_{ij} = -B_i^2\left(Z_{gi} + Z_{ij}\right) - |B_i||B_j|\left(M_{ij}\sin\beta_{ij} - Z_{ij}\cos\beta_{ij}\right) \end{cases} \tag{57}$$

where $\beta_{ij} = \beta_i - \beta_j$ is the voltage phase difference between the $i$ and $j$ nodes ($1 \le i, j \le 14$); $Q_i$ and $P_i$ illustrate the reactive power injection and real power injection; $|B|$ denote the voltage amplitude; $Z_{ij}$ and $M_{ij}$ illustrate the susceptance and conductance; $Q_{ij}$ and $P_{ij}$ illustrate the reactive power flow and real power flow; $M_{gi}$ and $Z_{gi}$ illustrate the conductance and susceptance of the Shunt at bus $i$, respectively.

Note that the actual measurements obtained in the system of equations (67) include noise. Suppose that a sensor network, as shown in Figure 5, is deployed, then the measurements of each node are calculated as follows:

$$\mathbf{v}_i^{(t)} = \mathbf{T}_1 + \mathbf{r}_i^{(t)} \quad ; i = 1,3,5,7,9 \tag{58}$$

$$\mathbf{v}_i^{(t)} = \mathbf{T}_2 + \mathbf{r}_i^{(t)} \quad ; i = 2,4,6,8,10 \tag{59}$$

where $\mathbf{T}_1$ and $\mathbf{T}_2$ are given in Table III.

TABLE III
MEASUREMENT INFORMATION IN $\mathbf{T}_1$ AND $\mathbf{T}_2$

| | |
|---|---|
| $\mathbf{T}_1$ | $B_1$, $\beta_1$, $Q_2$, $Q_4$, $Q_6$, $Q_8$, $Q_{10}$, $Q_{12}$, $Q_{14}$, $P_2$, $P_4$, $P_6$, $P_8$, $P_{10}$, $P_{12}$, $P_{14}$, $Q_{1-5}$, $Q_{2-4}$, $Q_{3-4}$, $Q_{4-7}$, $Q_{5-6}$, $Q_{6-12}$, $Q_{7-8}$, $Q_{9-10}$, $Q_{10-11}$, $Q_{13-14}$, $P_{1-5}$, $P_{2-4}$, $P_{3-4}$, $P_{4-7}$, $P_{5-6}$, $P_{6-12}$, $P_{7-8}$, $P_{9-10}$, $P_{10-11}$, $P_{13-14}$ |
| $\mathbf{T}_2$ | $B_1$, $\beta_1$, $Q_4$, $Q_5$, $Q_6$, $Q_8$, $Q_{10}$, $Q_{11}$, $Q_{12}$, $Q_{14}$, $P_4$, $P_5$, $P_6$, $P_8$, $P_{10}$, $P_{11}$, $P_{12}$, $P_{14}$, $Q_{1-2}$, $Q_{1-5}$, $Q_{2-3}$, $Q_{2-4}$, $Q_{2-5}$, $Q_{4-7}$, $Q_{6-11}$, $Q_{6-13}$, $Q_{12-13}$, $P_{1-2}$, $P_{1-2}$, $P_{1-5}$, $P_{2-3}$, $P_{2-4}$, $P_{2-5}$, $P_{4-7}$, $P_{6-11}$, $P_{6-13}$, $P_{12-13}$ |

*Scenario 1:* Investigation of the influence of $r_{mG}$ noise

In this scenario, the power system is affected by measurement noise $r_{mG}$ considered, in which the results of estimating voltage magnitude (V-M) and voltage angle (V-A) are illustrated in Figures 7 and 8. From the simulation results, it is clear that the proposed algorithm performs best in terms of accuracy and stability. The compensation of two different kernels and bandwidth flexibility allows AMKMMC-RDEKF to achieve outstanding performance. Specifically, in voltage magnitude estimation, it achieves 16%, 27%, 55%, and 70% higher accuracy than MKMMC-DEKF, MMC-DEKF, MCC-DEKF, and DEKF, respectively. At the same time, the obtained results once again confirm the recently published results, that is, in the corentropy family, MC has better efficiency and MKC has the best performance.

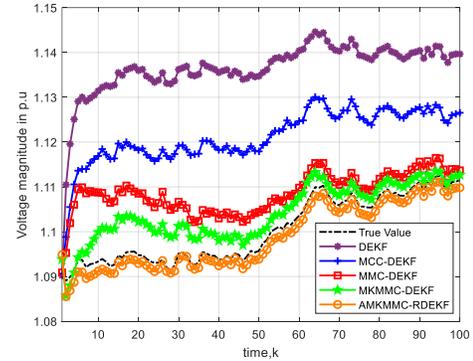

**Fig. 7.** Estimation of V-M for scenario 1.

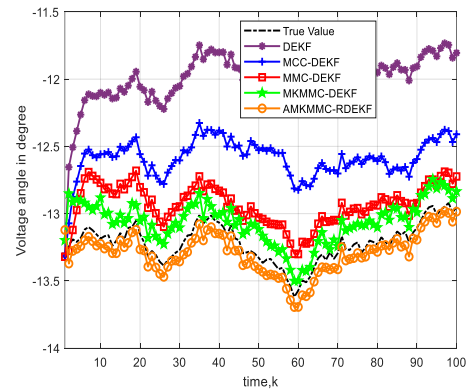

**Fig. 8.** Estimation of V-A for scenario 1.

***Scenario 2:*** Sudden load change

In Scenario 2, the ability of the proposed algorithm to handle non-Gaussian noise and maintain performance under complex conditions is evaluated. At the 40th time step, a sudden load change causes the voltage magnitude at bus-8 to drop by 15%. Simultaneously, $r_{mG}$ noise similar to that in Scenario 1 continues to affect the system. The estimated voltage magnitude and angle are shown in Figures 9 and 10. Additionally, by combining Scenarios 1 and 2, the average RMSE (ARMSE) results are presented in Table IV.

TABLE IV
ARMSE OF POWER SYSTEM STATE ESTIMATION

| ARMSE / Algorithm | Scenario 1 | | Scenario 2 | |
|---|---|---|---|---|
| | V-M | V-A | V-M | V-A |
| DEKF | 0.2028 | 0.0845 | 0.1989 | 0.0857 |
| MCC-DEKF | 0.1491 | 0.0722 | 0.1378 | 0.0668 |
| MMC-DEKF | 0.0920 | 0.0624 | 0.1109 | 0.0486 |
| MKMMC-DEKF | 0.0808 | 0.0581 | 0.0942 | 0.0408 |
| AMKMMC-RDEKF | **0.0676** | **0.0460** | **0.0744** | **0.0348** |

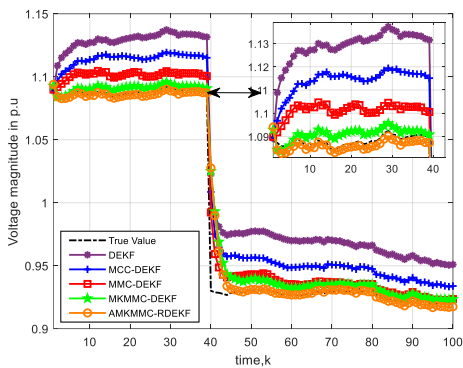

**Fig. 9.** Estimation of V-M for scenario 2.

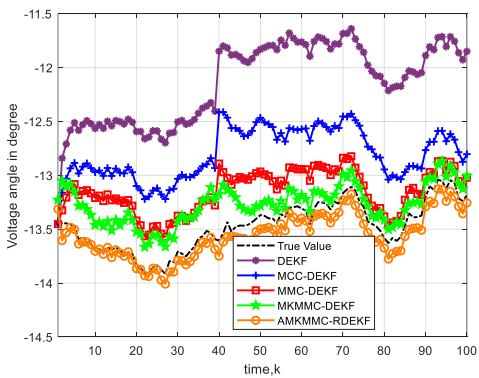

**Fig. 10.** Estimation of V-A for scenario 2.

The results demonstrate that the proposed AMKMMC-RDEKF algorithm delivers the most stable and accurate performance. Notably, during the abrupt change, it remains stable, while other algorithms exhibit abnormal estimates and require more time to recover. These findings confirm that AMKMMC-RDEKF satisfies the stringent demands of real-world applications.

## B. Land vehicle state estimation

The accuracy and stability of the proposed algorithm are further validated through land vehicle position and velocity estimation, where the state and measurement equations are defined as follows [45]:

$$\mathbf{u}_i = \begin{bmatrix} 1 & 0 & \Delta T & 0 \\ 0 & 1 & 0 & \Delta T \\ 0 & 0 & 1 & 0 \\ 0 & 0 & 0 & 1 \end{bmatrix} \mathbf{u}_{i-1} + \mathbf{q}_{i-1} \tag{60}$$

$$\mathbf{v}_i = \begin{bmatrix} -1 & 0 & -1 & 0 \\ 0 & -1 & 0 & -1 \end{bmatrix} \mathbf{v}_{i-1} + \mathbf{r}_i \tag{61}$$

where $\mathbf{u}_i = \begin{bmatrix} \mathbf{u}_{1,i} & \mathbf{u}_{2,i} & \mathbf{u}_{3,i} & \mathbf{u}_{4,i} \end{bmatrix}^T$ is a state vector (including north position, east position, north velocity, and east velocity); $\Delta T = 0.3$ : time interval.

Furthermore, the error covariance matrix $\mathbf{P}_i^{1|0}$, the prior estimate $\hat{\mathbf{u}}_i^{(0|0)}$, and the real state $\mathbf{u}_i^{(0|0)}$ are provided by:

$$\mathbf{P}_i^{1|0} = diag \begin{bmatrix} 900 & 900 & 4 & 4 \end{bmatrix} \tag{62}$$

$$\hat{\mathbf{u}}_i^{(0|0)} = \begin{bmatrix} 1 & 1 & 1 & 1 \end{bmatrix}^T \tag{63}$$

$$\mathbf{u}_i^{(0|0)} = \begin{bmatrix} 0 & 10 & \tan(\pi/3) & 10 \end{bmatrix}^T \tag{64}$$

***Scenario 3:*** Investigation of the influence of $r_{Ray}$ noise

In this scenario, the vehicle's position and velocity estimation process is affected by $r_{Ray}$ noise, where the estimation results are illustrated in Figures 11 and 12. From the simulation results, we observed that AMKMMC-RDEKF has the highest stability and accuracy. When faced with non-Gaussian noise $r_{Ray}$, algorithms developed based on the MKC criterion have shown better performance. The enhanced robustness of MKMMC over existing criteria, combined with the adaptive process for determining free coefficients, enables the proposed algorithm to maintain superior performance, even when the target of estimation changes.

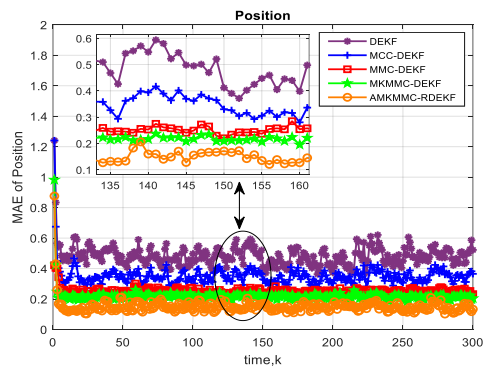

**Fig.11.** MAE of the position estimate for scenario 3.

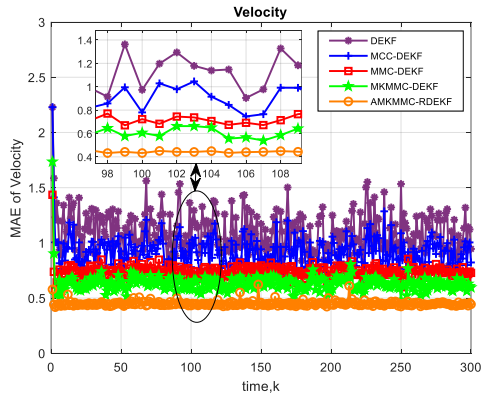

**Fig. 12.** MAE of the velocity estimate for scenario 3.

***Scenario 4:*** Packet loss

In this scenario, suppose that at the 100th test time, under the influence of $r_{mG}$ noise, the data transmission process lost 30% of packets [46,47,48]. The position and velocity estimation results are illustrated in Figures 13 and 14.

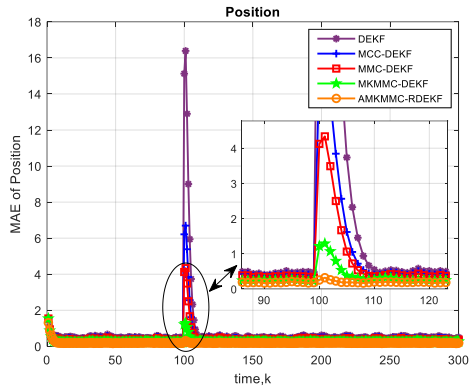

**Fig. 13.** MAE of the position estimate for scenario 4.

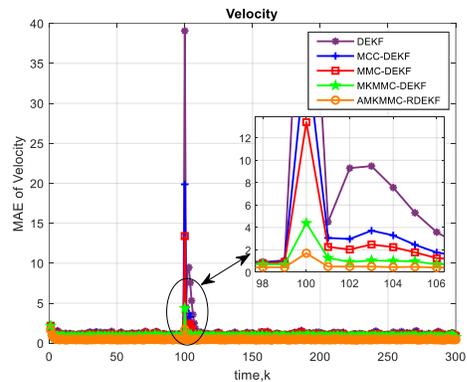

**Fig. 14.** MAE of the velocity estimate for scenario 4.

Considering the obtained results, it is evident that the distributed estimation algorithm we proposed has achieved the best stability and robustness. Specifically, at the time of packet loss, existing algorithms such as DEKF, MCC-DEKF, and MMC-DEKF exhibit large fluctuations and require a considerable amount of time to reach a stable estimation state, which does not meet the current requirements.

In particular, DEKF, which is based on the Gaussian assumption, exhibits substantial errors, consistent with the research results in [46]. In contrast, MKMMC-DEKF and its adaptive robust version AMKMMC-RDEKF consistently maintain superior accuracy. This is achieved by combining two different kernels with heavy-tailed features and optimized coefficients, which enhances the ability to model and handle unusually complex data. This study once again confirms the advantages of using multiple kernels, which have been analyzed and confirmed in studies [21,22,28,29].

### C. The influence of the number of kernels

Similar to existing studies based on multi-kernel, the MKMMC optimization criterion proposed in this study also uses two kernels. To answer the question: how many kernels will achieve the best estimation? A performance comparison experiment will be conducted between MKMMC with two, three, and four kernels. Specifically, the three-kernel and four-kernel MKMMC criteria are configured with (two Cauchy, one Student's *t*) and (two Cauchy, two Student's *t*) kernels, respectively. Under the same execution conditions as scenario 1, the estimation results are provided in Figures 15 and 16.

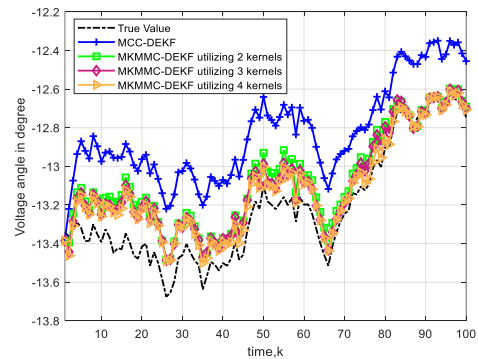

**Fig. 15.** Comparison of voltage angle estimates under different numbers of kernels

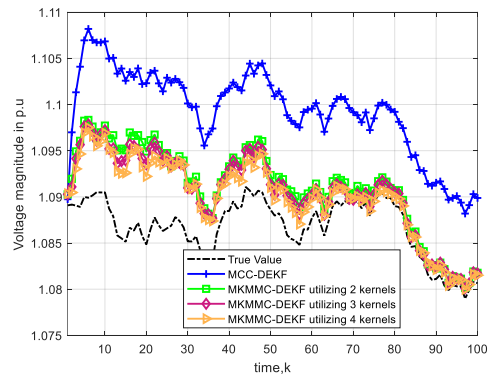

**Fig. 16.** Comparison of voltage magnitude estimates under different numbers of kernels

Observing the obtained results, it can be seen that the multi-kernel-based estimation algorithms have higher accuracy than those using a single kernel. However, comparing the algorithms using multiple kernels with each other, it can be seen that the difference is not large. On the other hand, increasing the number of kernels will face the challenge of increasing the number of coefficients to be selected and the computational complexity.

TABLE VI  CONVERGENCE SPEED

| Algorithm | ARMSE of position | | | ARMSE of velocity | | |
|---|---|---|---|---|---|---|
| | DEKF | MMC-DEKF | AMKMMC-RDEKF | DEKF | MMC-DEKF | AMKMMC-RDEKF |
| L=1 | 1.2518 | 0.8724 | 0.7012 | 2.3547 | 1.7236 | 1.3825 |
| L=2 | 0.8035 | 0.5421 | 0.4318 | 1.7562 | 1.2648 | 1.0237 |
| L=3 | 0.6329 | 0.4725 | 0.3614 | 1.3584 | 1.0243 | 0.8936 |
| L=5 | 0.6027 | 0.4118 | 0.3566 | 1.2839 | 0.9921 | 0.8893 |
| L=8 | 0.5876 | 0.3962 | 0.3479 | 1.2942 | 0.9835 | 0.8754 |
| L=10 | 0.5768 | 0.3853 | 0.3415 | 1.2731 | 0.9824 | 0.8706 |

For the problem of coefficients to be selected, if not well determined, the performance will be worse than that of algorithms using fewer kernels. Considering all aspects, the choice of using two kernels is an optimal choice. In addition, to realize the comparison of computational complexity, the single-step computation time in state estimation of the Land vehicle and the power system of the proposed algorithm with the existing algorithms is provided in Table V. In which the estimation condition is the same as scenarios 1 and 3. Considering the obtained results, it can be seen that the proposed MKMMC-DEKF is only equivalent to MMC-DEKF, in which the increase is insignificant. However, the ability to handle non-Gaussian noise, especially heavy-tailed noise, is better. With the great advances of current hardware devices, the small increase in time can be completely overcome. Therefore, the proposed algorithm is completely suitable for practical implementation.

TABLE V

COMPARE RUNNING TIMES

| Algorithms <br> Objects | DEKF | MCC-DEKF | MMC-DEKF | MKMMC-DEKF |
|---|---|---|---|---|
| Land vehicle | 0.0034 | 0.0052 | 0.0064 | 0.0067 |
| Power system | 0.0066 | 0.0081 | 0.0092 | 0.0096 |

### D. Convergence speed analysis

Table VI presents the ARMSEs of position and velocity at node-6 in Scenario 3 as a function of the consensus iteration number $L$. It is evident that the proposed AMKMMC-RDEKF achieves the lowest estimation error across all values of $L$, demonstrating superior robustness against outliers. More importantly, the algorithm exhibits rapid convergence: the ARMSE stabilizes after only $L = 3$ consensus iterations and remains unchanged for $L \geq 3$. This indicates that the proposed method requires minimal communication overhead to reach near-optimal performance, making it highly suitable for resource-constrained sensor networks. In contrast, both DEKF and MMC-DEKF show slower convergence and higher steady-state errors, particularly in velocity estimation, where DEKF's ARMSE remains above 1.27 even at $L = 10$. The fast convergence of AMKMMC-RDEKF can be attributed to two key factors: (i) the adaptive multi-kernel mixture (Student's $t$–Cauchy with non-zero means), which accurately models the heavy-tailed and asymmetric error distribution, and (ii) the closed-loop parameter adaptation mechanism, which dynamically aligns kernel centers and bandwidths with the empirical error statistics. These results confirm that the proposed algorithm not only improves estimation accuracy but also maintains computational and communication efficiency in distributed non-Gaussian environments.

### VII. CONCLUSION

In this paper, a novel MKMMC criterion has been developed to enhance the measurement between two random variables. Building upon this, two distributed filtering algorithms, MKMMC-DEKF and its adaptive robust version, AMKMMC-RDEKF, have been proposed. These algorithms design a kernel function formed by a mixture of two Student's $t$-Cauchy distributions with flexible mean values. Furthermore, the integration of an adaptive mechanism for selecting free coefficients and setting a measurement threshold significantly contributes to the improved stability and accuracy of AMKMMC-RDEKF. Simulation results on power system and land vehicle state estimation under complex conditions demonstrate the superior performance of the proposed methods.

### APPENDIX A

The first, Eq.(32) can be rewritten as:

$$\hat{\mathbf{P}}_k^{(t|t)} = \hat{\mathbf{P}}_k^{(t|t-1)} - \hat{\mathbf{P}}_k^{(t|t-1)} \left(\mathbf{H}_k^{(t|t)}\right)^T \left(\mathbf{K}_k^{(t|t)}\right)^T -$$
$$- \mathbf{K}_k^{(t|t)} \mathbf{H}_k^{(t|t)} \hat{\mathbf{P}}_k^{(t|t-1)} + \mathbf{K}_k^{(t|t)} \left[\mathbf{H}_k^{(t|t)} \hat{\mathbf{P}}_k^{(t|t-1)} \left(\mathbf{H}_k^{(t|t)}\right)^T + \mathbf{R}_k\right] \left(\mathbf{K}_k^{(t|t)}\right)^T$$
$$(65)$$

According to Eq.(29) and Eq.(65) can be approximated as

$$\hat{\mathbf{P}}_k^{(t|t)} \approx \hat{\mathbf{P}}_k^{(t|t-1)} - \hat{\mathbf{P}}_k^{(t|t-1)} \left(\mathbf{H}_k^{(t|t)}\right)^T \left(\mathbf{K}_k^{(t|t)}\right)^T -$$
$$- \mathbf{K}_k^{(t|t)} \mathbf{H}_k^{(t|t)} \hat{\mathbf{P}}_k^{(t|t-1)} + \hat{\mathbf{P}}_k^{(t|t-1)} \left(\mathbf{H}_k^{(t|t)}\right)^T \left(\mathbf{K}_k^{(t|t)}\right)^T$$
$$= \hat{\mathbf{P}}_k^{(t|t-1)} - \mathbf{K}_k^{(t|t)} \mathbf{H}_k^{(t|t)} \hat{\mathbf{P}}_k^{(t|t-1)}$$
$$= \hat{\mathbf{P}}_k^{(t|t-1)} - \hat{\mathbf{P}}_k^{(t|t-1)} \left(\mathbf{H}_k^{(t|t)}\right)^T \left[\mathbf{R}_k^{-1} + \mathbf{H}_k^{(t|t)} \hat{\mathbf{P}}_k^{(t|t-1)} \left(\mathbf{H}_k^{(t|t)}\right)^T\right]^{-1}$$
$$\times \mathbf{H}_k^{(t|t)} \hat{\mathbf{P}}_k^{(t|t-1)}$$
$$(66)$$

The matrix inverse lemma applied, Eq.(66) can be presented as

$$\hat{\mathbf{P}}_k^{(t|t)} = \left( \left( \hat{\mathbf{P}}_k^{(t|t-1)} \right)^{-1} + \left( \mathbf{H}_k^{(t|t)} \right)^T \mathbf{R}_k \mathbf{H}_k^{(t|t)} \right)^{-1} \quad (67)$$

Furthermore, the gain matrix $\mathbf{K}_k^{(t|t)}$ can be illustrated as:

$$\mathbf{K}_k^{(t|t)} = \hat{\mathbf{P}}_k^{(t|t)} \left( \mathbf{H}_k^{(t|t)} \right)^T \mathbf{R}_k \quad (68)$$

Then $\hat{\mathbf{u}}_k^{(t|t)}$ can be calculated as:

$$\begin{aligned}
\hat{\mathbf{u}}_k^{(t|t)} &= \hat{\mathbf{u}}_k^{(t|t-1)} + \hat{\mathbf{P}}_k^{(t|t)} \left( \mathbf{H}_k^{(t|t)} \right)^T \mathbf{R}_k \left( \hat{\mathbf{v}}_k^{(t|t)} - \mathbf{H}_k^{(t|t)} \hat{\mathbf{u}}_k^{(t|t-1)} \right) \\
&= \hat{\mathbf{P}}_k^{(t|t)} \left[ \left( \hat{\mathbf{P}}_k^{(t|t)} \right)^{-1} \hat{\mathbf{u}}_k^{(t|t-1)} + \left( \mathbf{H}_k^{(t|t)} \right)^T \mathbf{R}_k \left( \hat{\mathbf{v}}_k^{(t|t)} - \mathbf{H}_k^{(t|t)} \hat{\mathbf{u}}_k^{(t|t-1)} \right) \right] \\
&= \hat{\mathbf{P}}_k^{(t|t)} \left[ \begin{array}{l} \left( \left( \hat{\mathbf{P}}_k^{(t|t)} \right)^{-1} + \left( \mathbf{H}_k^{(t|t)} \right)^T \mathbf{R}_k \mathbf{H}_k^{(t|t)} \right) \hat{\mathbf{u}}_k^{(t|t-1)} \\ + \left( \mathbf{H}_k^{(t|t)} \right)^T \mathbf{R}_k \left( \hat{\mathbf{v}}_k^{(t|t)} - \mathbf{H}_k^{(t|t)} \hat{\mathbf{u}}_k^{(t|t-1)} \right) \end{array} \right] \\
&= \hat{\mathbf{P}}_k^{(t|t)} \left[ \left( \hat{\mathbf{P}}_k^{(t|t)} \right)^{-1} \hat{\mathbf{u}}_k^{(t|t-1)} \right) + \left( \mathbf{H}_k^{(t|t)} \right)^T \mathbf{R}_k \hat{\mathbf{v}}_k^{(t|t)} \right]
\end{aligned} \quad (69)$$

## APPENDIX B

***Proof***: The induction matrix norm is compatible with the associated $\ell_p$-norm vector, so that:

$$\left\| \mathbf{g}(\mathbf{u}) \right\|_1 = \left\| \left[ \mathbf{Y} + \gamma' \mathbf{I} \right]^{-1} \mathbf{F} \right\|_1 \leq \left\| \left[ \mathbf{Y} + \gamma' \mathbf{I} \right]^{-1} \right\|_1 \left\| \mathbf{F} \right\|_1 \quad (70)$$

with $\left\| \cdot \right\|_1$: denotes the 1-norm, commonly known as the column-sum norm-defined as the maximum absolute column sum of the matrix. According to the matrix theory, obtained:

$$\left\| \left[ \mathbf{Y} + \gamma' \mathbf{I} \right]^{-1} \right\|_1 \leq \sqrt{m} \left\| \left[ \mathbf{Y} + \gamma' \mathbf{I} \right]^{-1} \right\|_2 = \sqrt{m} \theta_{\max} \left\| \left[ \mathbf{Y} + \gamma' \mathbf{I} \right]^{-1} \right\| \quad (71)$$

with $\left\| \cdot \right\|_2$: denotes the 2-norm. In addition, there are:

$$\begin{aligned}
\theta_{\max} \left[ \left[ \mathbf{Y} + \gamma' \mathbf{I} \right]^{-1} \right] &= \frac{1}{\theta_{\min} \left[ \mathbf{Y} + \gamma' \mathbf{I} \right]} \\
&= \frac{1}{\theta_{\min} \left[ \sum_{j=1}^N \sum_{i=1}^m \frac{\theta_i}{\eta_i^2} \kappa_{\eta_i} \left( e_j - a_i \right) \mathbf{\Gamma}_j^T \mathbf{\Gamma}_j \right] + \gamma'} \\
&\leq \frac{\eta^3 \sqrt{2\pi}}{\theta_{\min} (\delta) + \gamma''}
\end{aligned} \quad (72)$$

where $\gamma'' = \eta^3 \sqrt{2\pi} \gamma'$. At the same time, get:

$$\begin{aligned}
\left\| \mathbf{F} \right\|_1 &= \left\| \sum_{j=1}^N \sum_{i=1}^m \frac{\theta_i}{\gamma_i^2 \eta^2} \kappa_{\eta_i} \left( e_j - a_i \right) \left( t_j - a_i \right) \mathbf{\Gamma}_j^T \right\| \\
&\leq \frac{1}{\eta^3 \sqrt{2\pi}} \sum_{j=1}^N \sum_{i=1}^m \frac{\theta_i}{\gamma_i^3} \left| t_j - a_i \right| \left\| \mathbf{\Gamma}_j^T \right\|_1
\end{aligned} \quad (73)$$

Combining Eq.(70)-Eq.(73), received:

$$\left\| \mathbf{g}(\mathbf{u}) \right\|_1 \leq \frac{\Lambda \sqrt{m}}{\theta_{\min} (\delta) + \gamma''} = \varphi(\eta) \quad (74)$$

It can be observed that the function $\varphi(\eta)$ is a monotonically decreasing and continuous function of $\eta$ in $(0, \infty)$, and satisfying $\lim_{\eta \to 0} \varphi(\eta) = \infty$ and $\lim_{\eta \to \infty} \varphi(\eta) = \zeta$. Therefore, if $\phi > \zeta$, the function $\varphi(\eta) = \phi$ will have a unique solution $\eta^*$ in $(0, \infty)$, and $\eta > \eta^*$, will have $\varphi(\eta) \leq \Upsilon$. This completes the proof.

## APPENDIX C

***Proof***: By Theorem 1, received $\left\| \mathbf{g}(\mathbf{u}) \right\|_1 \leq \phi$. To prove $\left\| \nabla_{\mathbf{u}} \mathbf{g}(\mathbf{u}) \right\|_1 \leq \hbar$, it suffices to prove:

$$\left\| \frac{\partial}{\partial u_s} \mathbf{g}(\mathbf{u}) \right\|_1 \leq \hbar, \; \forall s \quad (75)$$

$$\begin{aligned}
\left\| \frac{\partial}{\partial u_s} \mathbf{g}(\mathbf{u}) \right\|_1 &= \left\| \frac{\partial}{\partial u_s} \left( \left[ \mathbf{Y} + \gamma' \mathbf{I} \right]^{-1} \mathbf{F} \right) \right\|_1 \\
&\leq \left\| \left[ \mathbf{Y} + \gamma' \mathbf{I} \right]^{-1} \right\|_1 \left\| \frac{\partial}{\partial u_s} \left[ \mathbf{Y} + \gamma' \mathbf{I} \right] \right\|_1 \left\| \mathbf{g}(\mathbf{u}) \right\|_1 + \\
&\quad + \left\| \left[ \mathbf{Y} + \gamma' \mathbf{I} \right]^{-1} \right\|_1 \left\| \frac{\partial}{\partial u_s} \mathbf{F} \right\|_1
\end{aligned} \quad (76)$$

It can be easily derived:

$$\begin{aligned}
&\left\| \frac{\partial}{\partial u_s} \left[ \mathbf{Y} + \gamma' \mathbf{I} \right] \right\|_1 = \\
&= \left\| \sum_{j=1}^N \sum_{i=1}^m \frac{\theta_i}{\gamma_i^4 \eta^4} \left( e_j - a_i \right) \mathbf{\Gamma}_{js} \kappa_\eta \left( e_j - a_i \right) \mathbf{\Gamma}_j \mathbf{\Gamma}_j^T \right\|_1 \\
&\leq \frac{1}{\eta^5 \sqrt{2\pi}} \sum_{j=1}^N \sum_{i=1}^m \frac{\theta_i}{\gamma_i^5} \left( \phi \left\| \mathbf{\Gamma}_j \right\|_1 + \left| t_j - a_i \right| \right) \left\| \mathbf{\Gamma}_j \right\|_1 \left\| \mathbf{\Gamma}_j^T \mathbf{\Gamma}_j \right\|_1
\end{aligned} \quad (77)$$

Similarly, received:

$$\begin{aligned}
&\left\| \frac{\partial}{\partial u_s} \mathbf{F} \right\|_1 \leq \\
&\leq \frac{1}{\eta^5 \sqrt{2\pi}} \sum_{j=1}^N \sum_{i=1}^m \frac{\theta_i}{\gamma_i^5} \left( \phi \left\| \mathbf{\Gamma}_j \right\|_1 + \left| t_j - a_i \right| \right) \left\| \mathbf{\Gamma}_j \right\|_1 \left\| \mathbf{\Gamma}_j^T \mathbf{\Gamma}_j \right\|_1
\end{aligned} \quad (78)$$

Then, combining Eq.(71), Eq.(72), Eq.(76)-Eq.(78), and $\left\| \mathbf{g}(\mathbf{u}) \right\|_1 \leq \phi$, obtained:

$$\left\| \frac{\partial}{\partial u_s} \mathbf{g}(\mathbf{u}) \right\|_1 \leq \chi(\eta) \quad (79)$$

It can be observed that the function $\chi(\eta)$ is a monotonically decreasing and continuous function of $\eta$ in $(0, \infty)$, and satisfying $\lim_{\eta \to 0} \chi(\eta) = \infty$ and $\lim_{\eta \to \infty} \chi(\eta) = 0$. Therefore, given $1 > \hbar > 0$, the equation $\chi(\eta) = \hbar$ has a unique solution $\eta^*$ in $(0, \infty)$, and $\eta > \eta^*$, obtained: $\chi(\eta) = \phi$. This completes the proof.

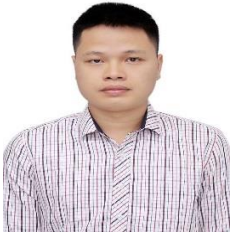

**Duc Viet Nguyen** received a master's degree in control engineering and automation from Le Quy Don University, Viet Nam, in 2019. He is working toward the Ph.D. degree in signal and information processing from the School of Electrical Engineering at Southwest Jiaotong University, Chengdu, China.

His current research interests include designing automatic control systems, state estimation, and adaptive filtering algorithms.

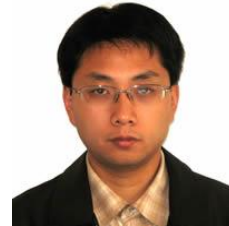

**Haiquan Zhao** (Senior Member, IEEE) received the B.S. degree in applied mathematics and the M.S. and Ph.D. degrees in signal and information processing from Southwest Jiaotong University, Chengdu, China, in 1998, 2005, and 2011, respectively.

Since 2012, he has been a Professor with the School of Electrical Engineering, Southwest Jiaotong University. From 2015 to 2016, he was a Visiting Scholar with the University of Florida, Gainesville, FL, USA. He is the author or coauthor of more than 280 international journal papers (SCI indexed) and owns 56 invention patents. His current research interests include information theoretical learning, adaptive filters, adaptive networks, active noise control, Kalman filters, machine learning, and artificial intelligence.

Dr. Zhao has won several provincial and ministerial awards and many best paper awards at international conferences or IEEE TRANSACTIONS. He has served as an Active Reviewer for several IEEE TRANSACTIONS, IET series, signal processing, and other international journals. He is currently a Handling Editor of Signal Processing, and also an Associate Editor for IEEE Transaction on Audio, Speech and Language Processing, IEEE TRANSACTIONS ON SYSTEMS, MAN AND CYBERNETICS: SYSTEM, IEEE SIGNAL PROCESSING LETTERS, IEEE SENSORS JOURNAL, and IEEE OPEN JOURNAL OF SIGNAL PROCESSING.

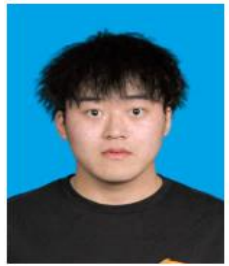

**Jinhui Hu** received the B.E. degree in electrical engineering and automation from Chang' An University, Xi'an, China, in 2022. He is currently working toward the Ph.D degree in signal and information processing from the School of Electrical Engineering, Southwest Jiaotong University, Chengdu, China.

His current research interests include state estimation and adaptive filtering algorithms.

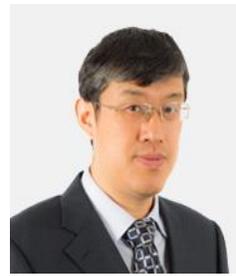

**Xiaoli Li** (Fellow, IEEE) is currently the department head and senior principal scientist with the Institute for Infocomm Research, A∗STAR, Singapore. He also serves as an adjunct full professor with the College of Computing and Data Science, Nanyang Technological University, Singapore. With a diverse range of research interests, Xiaoli focuses on cutting-edge areas such as AI, data mining, machine learning, and bioinformatics. His contributions to these fields are evident through his extensive publication record, boasting more than 350 peer-reviewed papers, and the recognition he has received, including over ten best paper awards. He has been serving as the editor-in-chief for Annual Review of Artificial Intelligence and an associate editor for prestigious journals like IEEE Transactions on Artificial Intelligence and Knowledge and Information Systems, as well as conference chairs and area chairs of leading AI, machine learning, and data science conferences, such as AAAI, IJCAI, ICLR, NeurIPS, KDD, ICDM,etc. Beyond academia, Xiaoli possesses extensive industry experience, where he has successfully spearheaded more than 10 R&D projects in collaboration with major industry players across diverse sectors, such as aerospace, telecom, insurance, and professional service companies. He is the fellow of Asia-Pacific Artificial Intelligence Association (AAIA). He has been recognized as one of the world's top 2% scientists in the AI domain by Stanford University and one of the top ranked computer scientists by Research.com.